\documentclass[epj,twocolumn]{webofc}
\usepackage[varg]{txfonts}   
\usepackage{upgreek}
\DeclareUnicodeCharacter{00A0}{ }
\begin{document}
\def\Pchi    {\ensuremath{\upchi}\xspace}         
\def\PJ      {\ensuremath{\mathrm{J}}\xspace}     
\def\Ppsi   {\ensuremath{\uppsi}\xspace}  
 \def\Pupsilon    {\ensuremath{\upsilon}\xspace}          
\def\jpsi     {\ensuremath{{\PJ\mskip -3mu/\mskip -2mu\Ppsi\mskip 2mu}}\xspace}     
\def\psitwos  {\ensuremath{\Ppsi{(2S)}}\xspace}
\def\chiczero {\ensuremath{\Pchi_{\cquark 0}}\xspace}
\newcommand{\lum}{\ensuremath{\mathcal{L}}}
\newcommand{\tev}{\ensuremath{\,\mathrm{TeV}}}
\newcommand{\gevc}{\ensuremath{\,\mathrm{GeV}\mskip -2mu/\mskip -1mu c}}
\newcommand{\mevc}{\ensuremath{\,\mathrm{MeV}\mskip -2mu/\mskip -1mu c}}
\newcommand{\gevcc}{\ensuremath{\,\mathrm{GeV}\mskip -2mu/\mskip -1mu c^2}}
\newcommand{\mevcc}{\ensuremath{\,\mathrm{MeV}\mskip -2mu/\mskip -1mu c^2}}
\newcommand{\pbinv}{\ensuremath{\,\mathrm{pb}^{-1}}}
\newcommand{\nbinv}{\ensuremath{\,\mathrm{nb}^{-1}}}
\newcommand{\bhad}{$b$-hadron}
\newcommand{\prompt}{\ensuremath{\mathrm{prompt}~\jpsi}}
\newcommand{\fromb}{\ensuremath{\jpsi~\mathrm{from}~b}}
\newcommand{\pt}{\ensuremath{p_{\rm T}}}
\newcommand{\microb}{\ensuremath{\,\upmu\mathrm{b}}}
\newcommand{\pp}{\ensuremath{\mbox{p-p}}}
\newcommand{\tpm}{$ & $\,\pm\,$ & $}
\newcommand{\upsmm}{\nobreak{\ensuremath{\varUpsilon\rightarrow \mu^+\mu^-}}}
\newcommand{\ups}{\ensuremath{\varUpsilon}}
\newcommand{\upsns}{\ensuremath{\varUpsilon(nS)}}
\newcommand{\ones}{\ensuremath{\varUpsilon(1S)}}
\newcommand{\twos}{\ensuremath{\varUpsilon(2S)}}
\newcommand{\threes}{\ensuremath{\varUpsilon(3S)}}
\newcommand{\ismm}{\ensuremath{\varUpsilon(nS)\rightarrow\mu^{+}\mu^{-}}}
\newcommand{\onesmm}{\ensuremath{\varUpsilon(1S)\rightarrow\mu^+\mu^-}}
\newcommand{\twosmm}{\ensuremath{\varUpsilon(2S)\rightarrow\mu^+\mu^-}}
\newcommand{\threesmm}{\ensuremath{\varUpsilon(3S)\rightarrow\mu^+\mu^-}}
\newcommand{\bmm}{\ensuremath{\mathcal{B}(\mu^+\mu^-)}}
\newcommand{\bmmis}{\ensuremath{\mathcal{B}(\varUpsilon(nS)\rightarrow\mu^+\mu^-)}}
\newcommand{\bmmones}{\ensuremath{\mathcal{B}(\varUpsilon(1S)\rightarrow\mu^+\mu^-)}}
\newcommand{\bmmtwos}{\ensuremath{\mathcal{B}(\varUpsilon(2S)\rightarrow\mu^+\mu^-)}}
\newcommand{\bmmthrees}{\ensuremath{\mathcal{B}(\varUpsilon(3S)\rightarrow\mu^+\mu^-)}}
\def\pythia     {\mbox{\textsc{Pythia}}\xspace}

%
%
%
%
\woctitle{LHCP 2013}
%
%
%
\title{Quarkonium Production at LHCb}
%
%

\author{Monica Pepe Altarelli\inst{1}
\fnsep\thanks
{\email{monica.pepe.altarelli@cern.ch}} 
\\  On behalf of the LHCb collaboration
}
\institute{CERN}

\abstract{%
I will review a selection of LHCb results on the production of heavy quarkonium states  in {\it pp} collisions, including recent results on  \jpsi\ and \upsns\ ($n=1,2,3$) production at $\sqrt{s}=8\tev$,  as well as  preliminary results on \jpsi\ production in proton-lead collisions at $\sqrt{s_{\rm NN}}=5\tev$.
}
\maketitle
\section{Introduction}
\label{intro}
Heavy quarkonium states are particularly interesting systems to test our understanding of strong interactions, both at perturbative and non-perturbative level, and are therefore the subject of intense theoretical and experimental studies. An~effective field theory, non-relativistic QCD (NRQCD)~\cite{CaswellLepage1986PL,PhysRevD.51.1125}, provides the foundation for much of the current theoretical work.  According to NRQCD, the production of heavy quarkonium factorises into two steps:  a heavy quark-antiquark pair is first created at short distances and subsequently evolves non-perturbatively into quarkonium at long distances. The NRQCD calculations include colour-singlet (CS) and colour-octet (CO) amplitudes (see~\cite{Brambilla:2010cs}  and references therein), which account for the probability of a heavy quark-antiquark pair in a particular colour state to evolve into a heavy quarkonium state. The CS model (CSM) was initially used to describe experimental data. However, it underestimates the observed cross-section for single \jpsi~production at high transverse momentum (\pt). To~resolve this discrepancy, the CO mechanism was introduced. More recent higher-order calculations to the CS predictions raise substantially the cross-sections at large \pt\ bringing them closer to the experimental data. However, none of these approaches can reproduce in a consistent way the available experimental results on both cross-section and polarisation (see~\cite{Brambilla:2010cs}  and references therein,~\cite{Lourenco}).

Heavy charmonium is also produced from $b$-hadron decays. This production mechanism can be distinguished from prompt quarkonium production by exploiting the $b$-hadron finite  decay time. In this case QCD predictions are based on the Fixed-Order-Next-to-Leading-Log (FONLL) formalism~\cite{FONLL,Cacciari}, which combines a Fixed-Order calculation with an all-order resummation of Leading and Next-to-Leading logarithms.

In the following, results from heavy quarkonium production at LHCb will be compared with various theoretical models to provide direct tests of the underlying
production mechanism. Furthermore,  preliminary results on \jpsi\ production in proton-lead collisions at $\sqrt{s_{\rm NN}}=5\tev$ will be reported.
\section{Production of \jpsi\ and \upsns\ mesons}
\label{sec-1}
\begin{figure}
\centering
\includegraphics[width=7cm,clip]{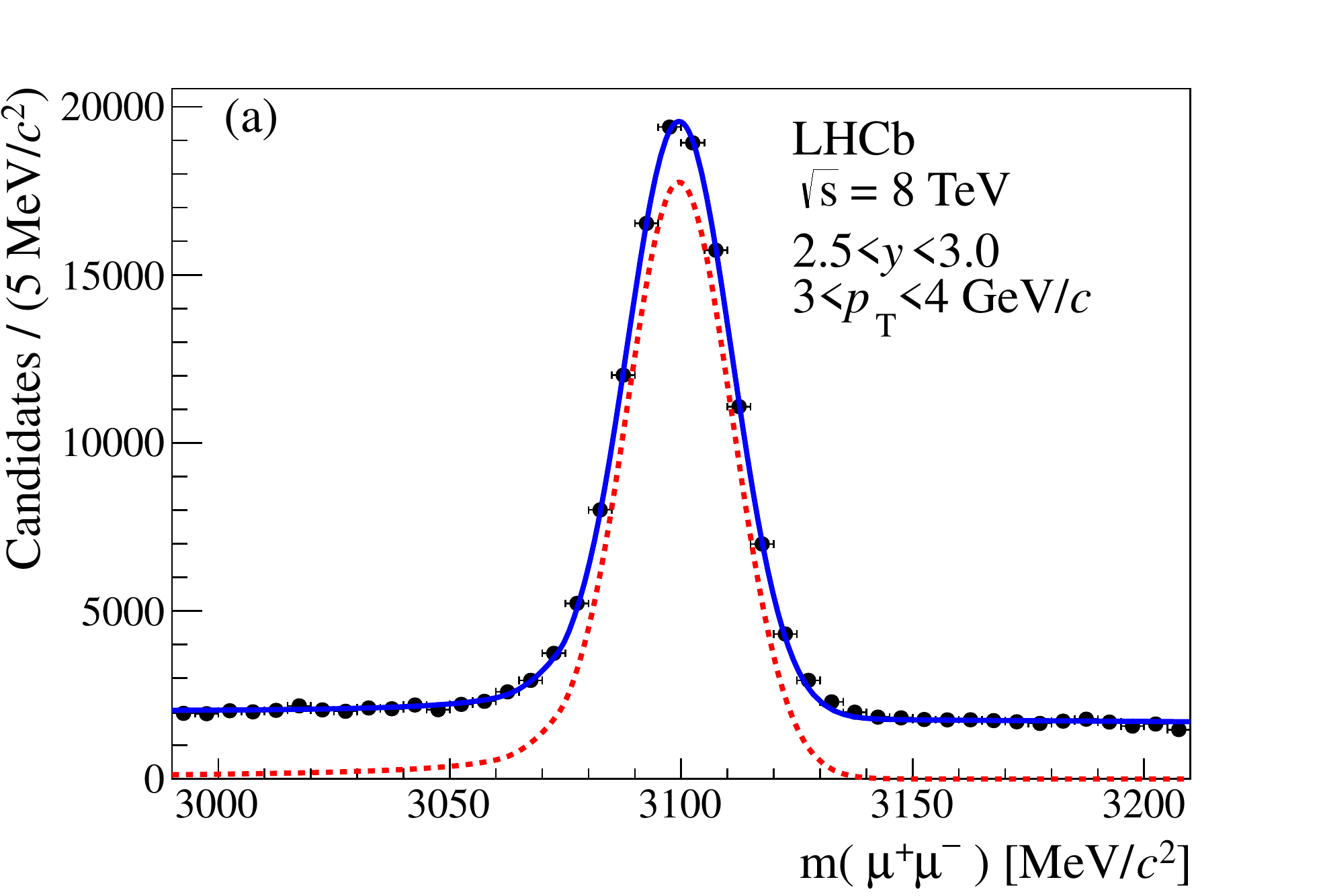}
\caption{\small  Dimuon invariant mass of the selected $\jpsi\to\mu^+\mu^-$ candidates in a selected bin in  $y$ and \pt\ ($2.5<y<3.0$ and $3<\pt<4\,\gevc$).  }
\label{fig-1}       
\end{figure}

\begin{figure}
\centering
\includegraphics[width=7cm,clip]{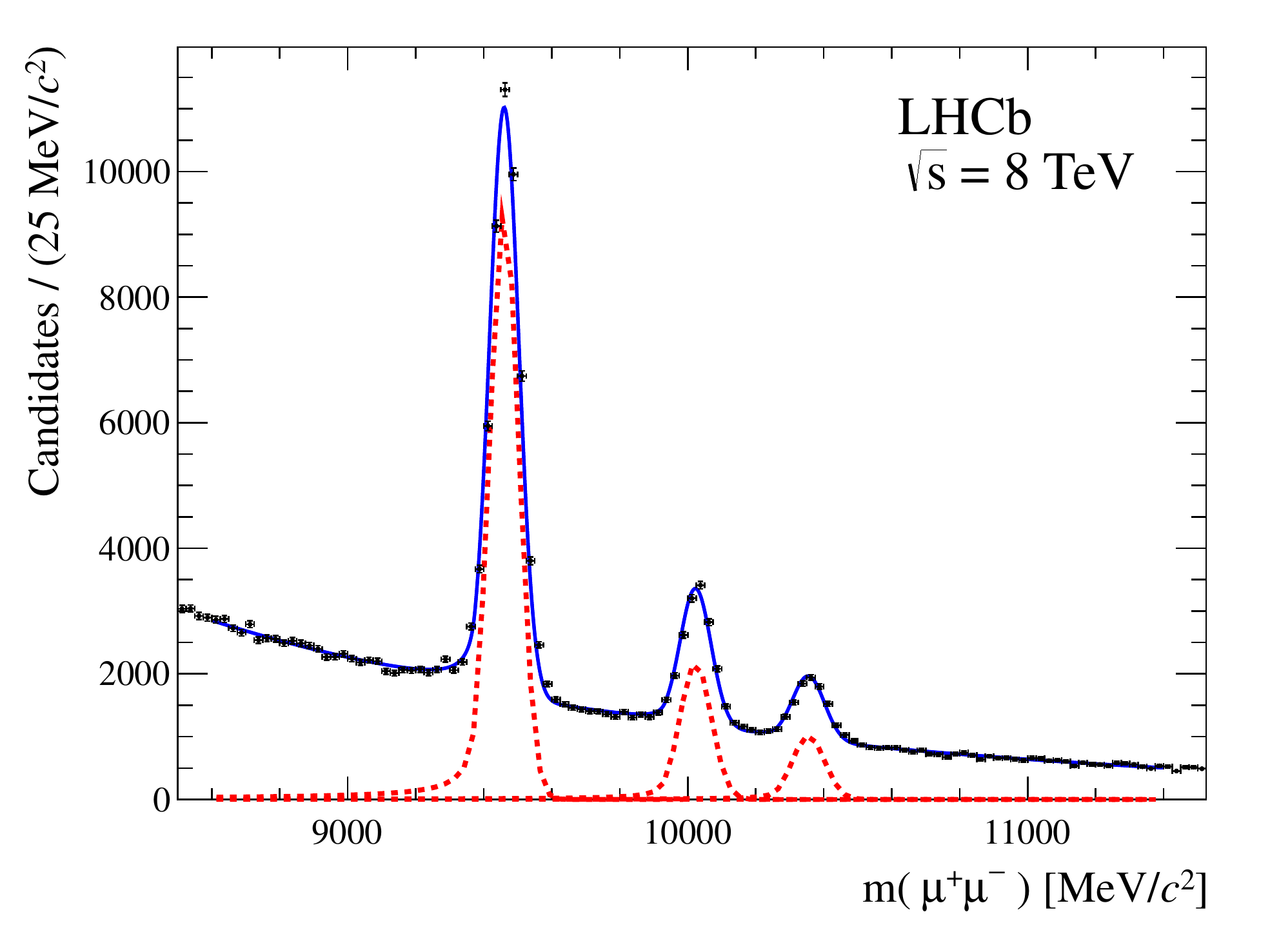}
\caption{\small Invariant mass distribution of the selected $\upsns\to\mu^+\mu^-$ candidates in the range $\pt<15\,\gevc$ and
$2.0<y<4.5$. The three peaks correspond to the $\ones$, $\twos$ and $\threes$ meson signals (from left to right).  }
\label{fig-2}       
\end{figure}

The differential production cross-sections of prompt \jpsi\ and \upsns\ mesons produced at the $pp$ collision point either directly or via feed-down from higher mass charmonium or bottomonium states, are measured  at $\sqrt{s}=8\tev$~\cite{Aaij:2013yaa} with the LHCb detector~\cite{Alves:2008zz}. The measurements are performed in the range of rapidity $2.0<y<4.5$ and $\pt<14\,\gevc$ for the \jpsi\ meson and $\pt<15\,\gevc$ for the \upsns, in bins of \pt\ and $y$ with bin sizes $\Delta \pt=1\gevc$ and $\Delta y=0.5$.
The fraction of \fromb\   (abbreviated as ``from $b$'' in the following) is also measured in the same fiducial region.
The \upsns\ meson analysis is based on a data sample, corresponding to an integrated luminosity of about $51~\pbinv$ of $pp$ interactions, while the analysis for the more abundant  \jpsi\ mesons is based on data, corresponding to an  integrated luminosity of about $18~\pbinv$. The  \jpsi\ and \upsns\ mesons are reconstructed through their decay into a pair of muons following the strategy described in Refs.~\cite{Aaij:2013yaa,LHCb-PAPER-2011-003,LHCB-PAPER-2011-036}. The resulting dimuon invariant mass distributions  are displayed in Figs.\ref{fig-1} and~\ref{fig-2} for the \jpsi\ and \upsns\ mesons. The invariant mass resolution is on average  $13.5\mevcc$ for the \jpsi\ and  $43\mevcc$ for the \ones\ meson.

The production is studied under the assumption of zero polarisation, and no corresponding systematic uncertainty is assigned
on the cross-section for this effect. This assumption is justified by the small polarisation measured for the \jpsi\ meson at $\sqrt{s}=7~\tev$ 
by LHCb~\cite{LHCb-PAPER-2013-008} and ALICE~\cite{AliceJpsiPola}, 
in a kinematic range similar to that used in this analysis, and for the
$\upsns$  by CMS~\cite{CMSUpsilonPola} 
at large $\pt$ and central rapidity.

\begin{figure}
\centering
\includegraphics[width=8cm,clip]{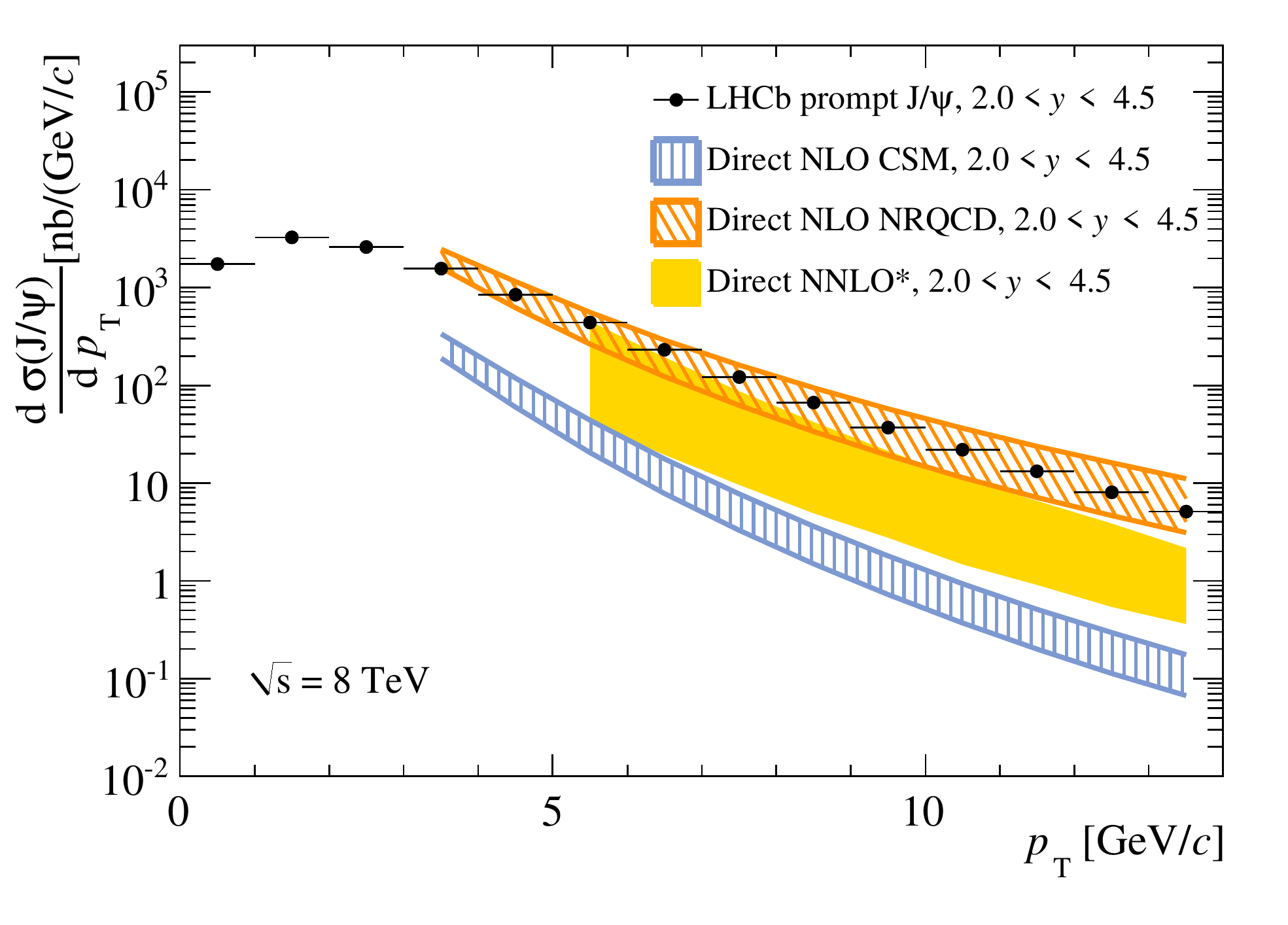}
\caption{\small Comparison of the differential cross-section for
   the production of prompt \jpsi\ meson (under the assumption of zero polarisation) as a function of
   \pt\  with direct production in an NLO NRQCD model~\cite{Butenschoen:2011yh, Butenschoen:2010rq}
   (orange diagonal shading), an NNLO* CSM~\cite{lansberg:2009} (solid yellow) and an
   NLO CSM~\cite{Campbell:2007ws}  (blue vertical shading). }
\label{fig-3}       
\end{figure}

The measured differential cross-sections for the production of prompt \jpsi\ mesons as a function of $\pt$ are compared in Fig.\ref{fig-3}  to three theoretical models: an NRQCD model at next-to-leading order (NLO)~\cite{Butenschoen:2011yh, Butenschoen:2010rq}, an NNLO* CSM~\cite{Artoisenet:2008,lansberg:2009} where the notation NNLO* indicates that the calculation  at next-to-next leading order is not complete and neglects part of the logarithmic terms and an NLO CSM~\cite{Campbell:2007ws}.
In these comparisons the predictions are for  direct \jpsi\ meson production, {\it i.e.}, they do not include feed-down from higher charmonium states,
whereas the experimental measurements do include feed-down. In particular, the contribution 
from \jpsi\ mesons produced in radiative $\rm\chi_c$ decays in the considered
fiducial range was measured to be at the level of $20\%$ at $\sqrt{s}= 7\,\tev$~\cite{LHCb:2012af}. Allowing for this contribution, both the NNLO* CSM and the NLO NRQCD models provide
reasonable descriptions of the experimental data. In contrast, the CSM at NLO
underestimates the cross-section by an order of magnitude.

%

\begin{figure}
\centering
\includegraphics[width=8cm,clip]{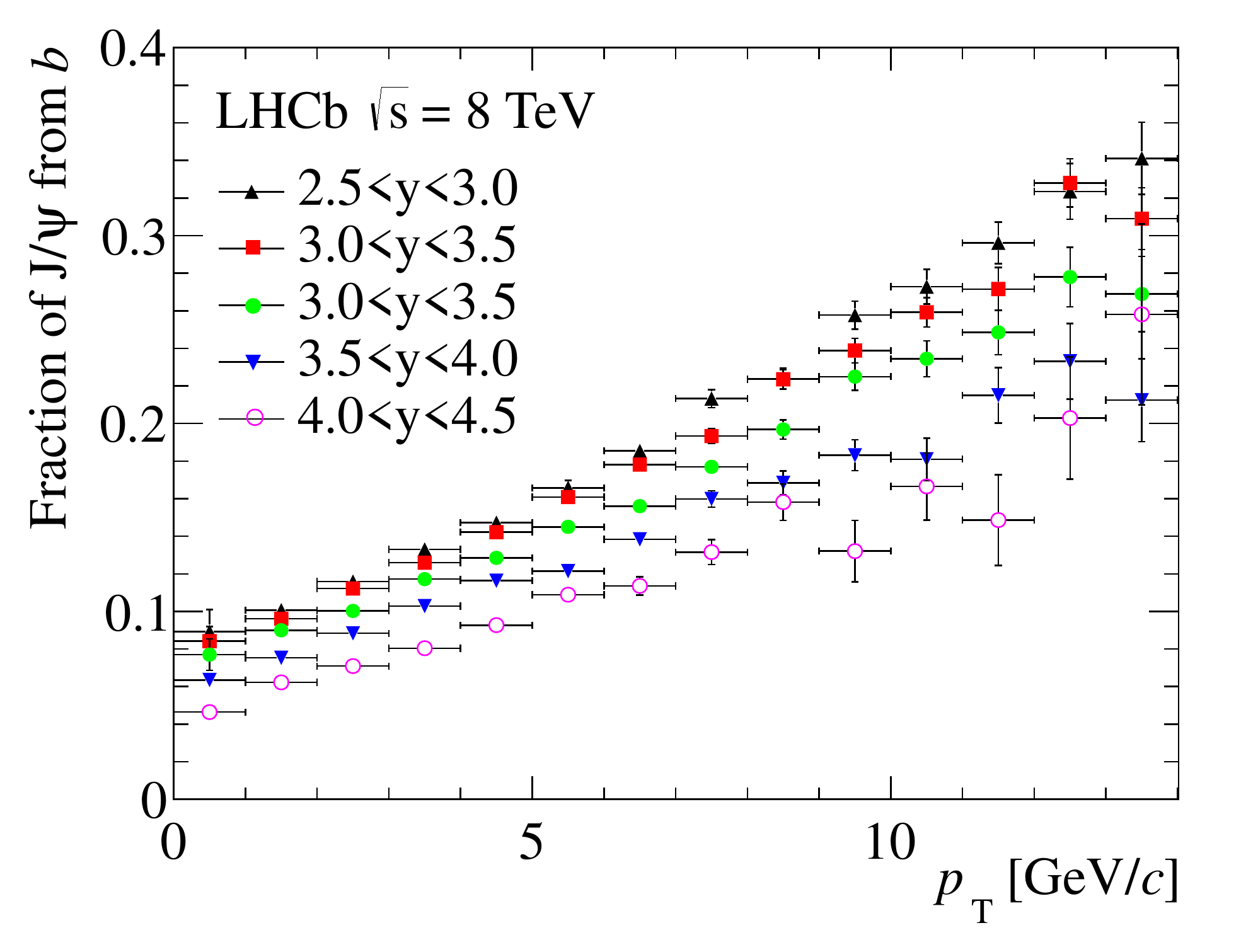}
\caption{\small  Fraction of \fromb\ as a function of the \jpsi\ transverse momentum, in bins of rapidity.  }
\label{fig-4}       
\end{figure}

\begin{figure}
\centering
\includegraphics[width=7cm,clip]{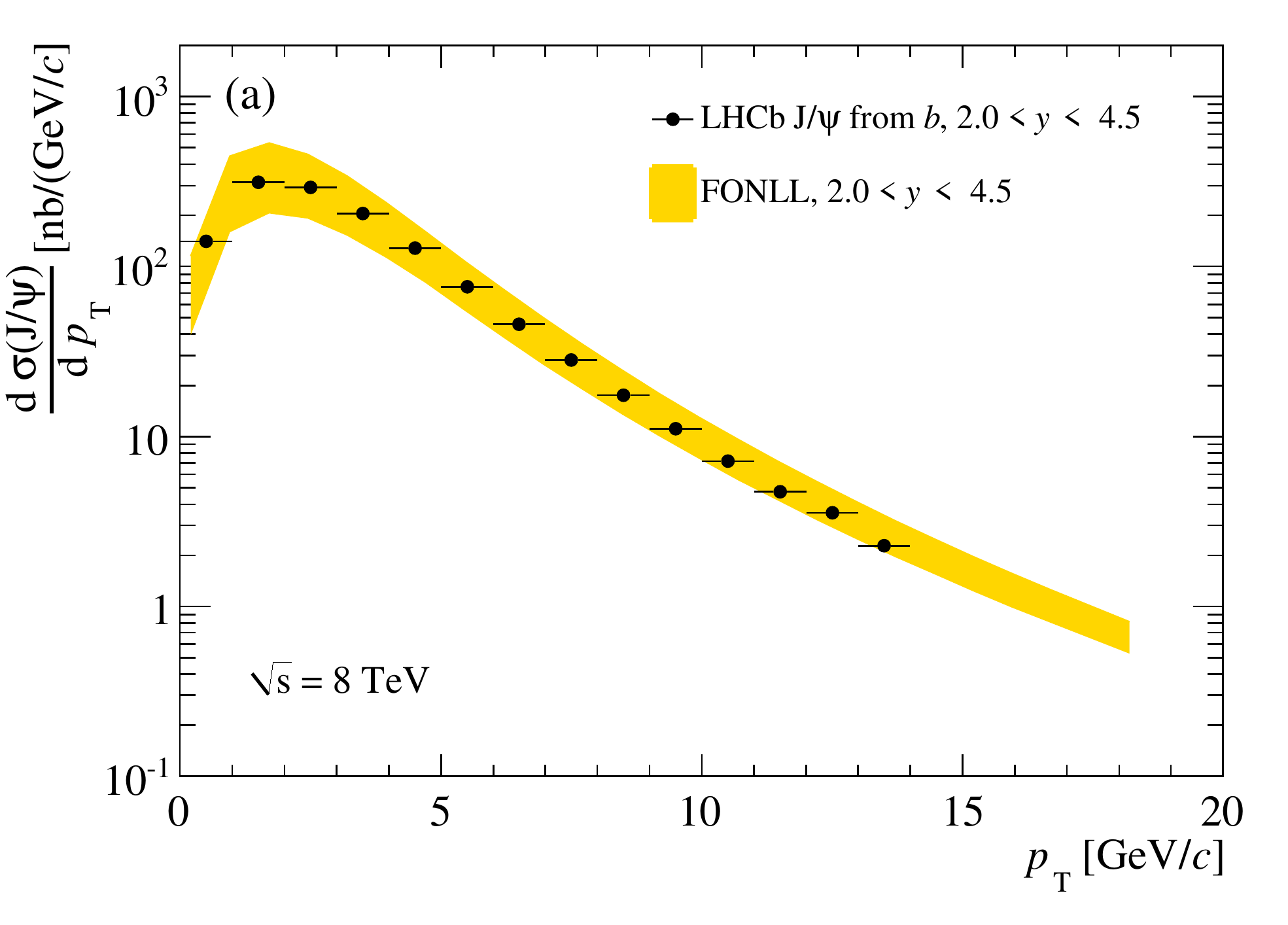}
\caption{\small  Differential production cross-section for \fromb\  as a function of \pt\ in the
  fiducial range $2.0<y<4.5$. The FONLL prediction~\cite{FONLL,Cacciari} is shown in yellow. }
\label{fig-5}       
\end{figure}

The fraction of \fromb\ is shown is Fig.\ref{fig-4} as a function of the \jpsi\ transverse momentum in bins of rapidity. This fraction is measured to be smaller at low \pt\ and for forward data.
In Fig.\ref{fig-5} the data for the differential production cross-section for \fromb\  as a function of \pt\  at $\sqrt{s} = 8\,\tev$  are compared to the FONLL predictions. Very good agreement is observed.
The integrated cross-section for \fromb\ production in the defined fiducial region is determined to be $\sigma\left(\fromb,\, \pt <14\,\gevc,\,2.0<y<4.5\right)  = 1.28 \pm 0.01\pm 0.11\microb$~\cite{Aaij:2013yaa}, where the first uncertainty is statistical and the second is systematic, in very good agreement with the FONLL prediction of  $1.34^{+0.63}_{-0.49}\,\upmu{\rm b}$~\cite{Cacciari}. 

\begin{figure}
\centering
\includegraphics[width=7cm,clip]{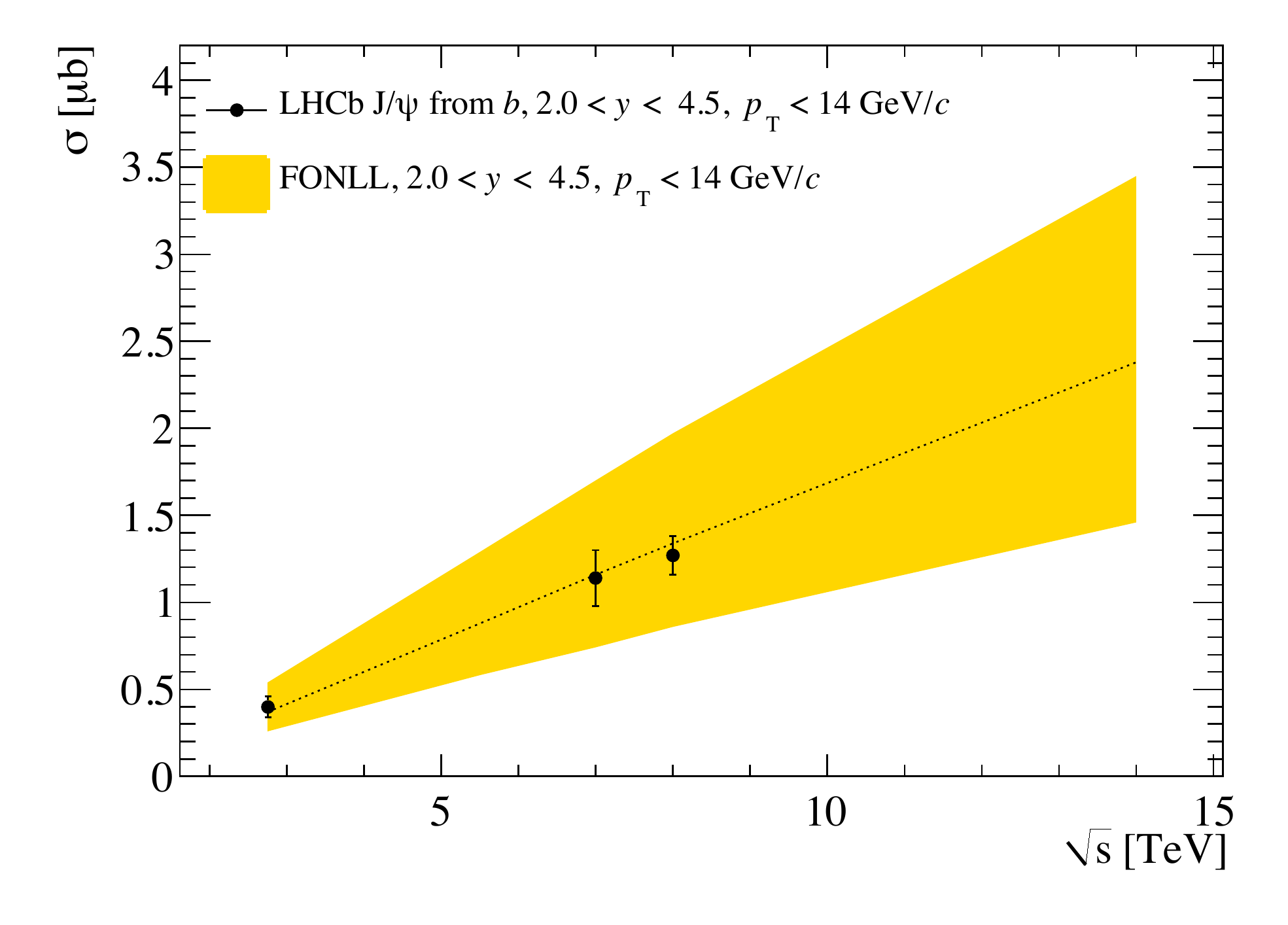}
\caption{\small Predictions based on the FONLL formalism~\cite{FONLL,Cacciari} for the production cross-section for \fromb\ in the fiducial range $0<\pt<14\,\gevc$ and $2.0<y<4.5$ (yellow band). The uncertainty includes contributions from the renormalisation scale, quark masses and the choice of PDF set.  The black dotted line shows the central value of the prediction. The points show the LHCb measurements at $\sqrt{s}=2.76$~\cite{LHCb-PAPER-2012-039}, 7~\cite{LHCb-PAPER-2011-003},  and 8$\,\tev$~\cite{Aaij:2013yaa}. }
\label{fig-6}
\end{figure}

The results on \jpsi\ production at $\sqrt{s}=8\tev$
provide an extra measurement, which complements those already published by LHCb at $\sqrt{s}=2.76$~\cite{LHCb-PAPER-2012-039} and  7~\tev\cite{LHCb-PAPER-2011-003}. Figure~\ref{fig-6} shows the results for the cross-section for \fromb\
at  $\sqrt{s}=2.76$, 7 and $8\,\tev$ as well as the FONLL prediction  as a function of centre-of-mass energy. Excellent agreement with the theoretical calculation is observed.

From the measurement of the production cross-section for \fromb\ in the fiducial region, using the LHCb Monte Carlo simulation based on \pythia to extrapolate to the full polar angle range (with an extrapolation factor of 5.4  from the measured cross-section to the full kinematic region)
and the inclusive $b{\to}\jpsi X$ branching fraction, one can derive the total $b\overline{b}$ production cross-section at $\sqrt{s}=8\tev$, \mbox{$\sigma(pp \to b\overline{b} X) = 298\pm 2 \pm 36\, \upmu{\rm b}$}, where the first uncertainty is statistical and the second is systematic. 

\begin{figure}
\centering
  \includegraphics[width=7cm,clip]{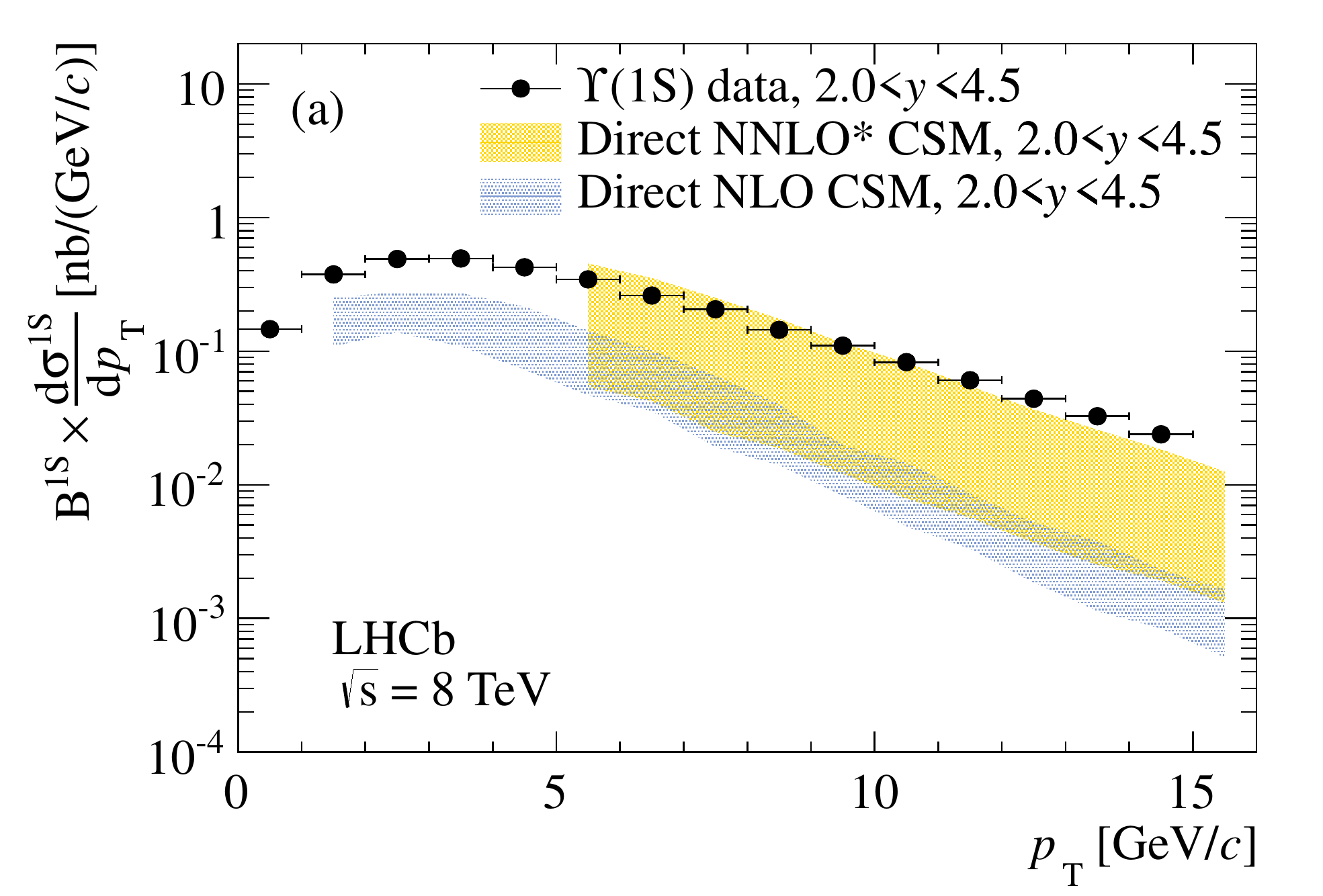}\\
  \includegraphics[width=7cm,clip]{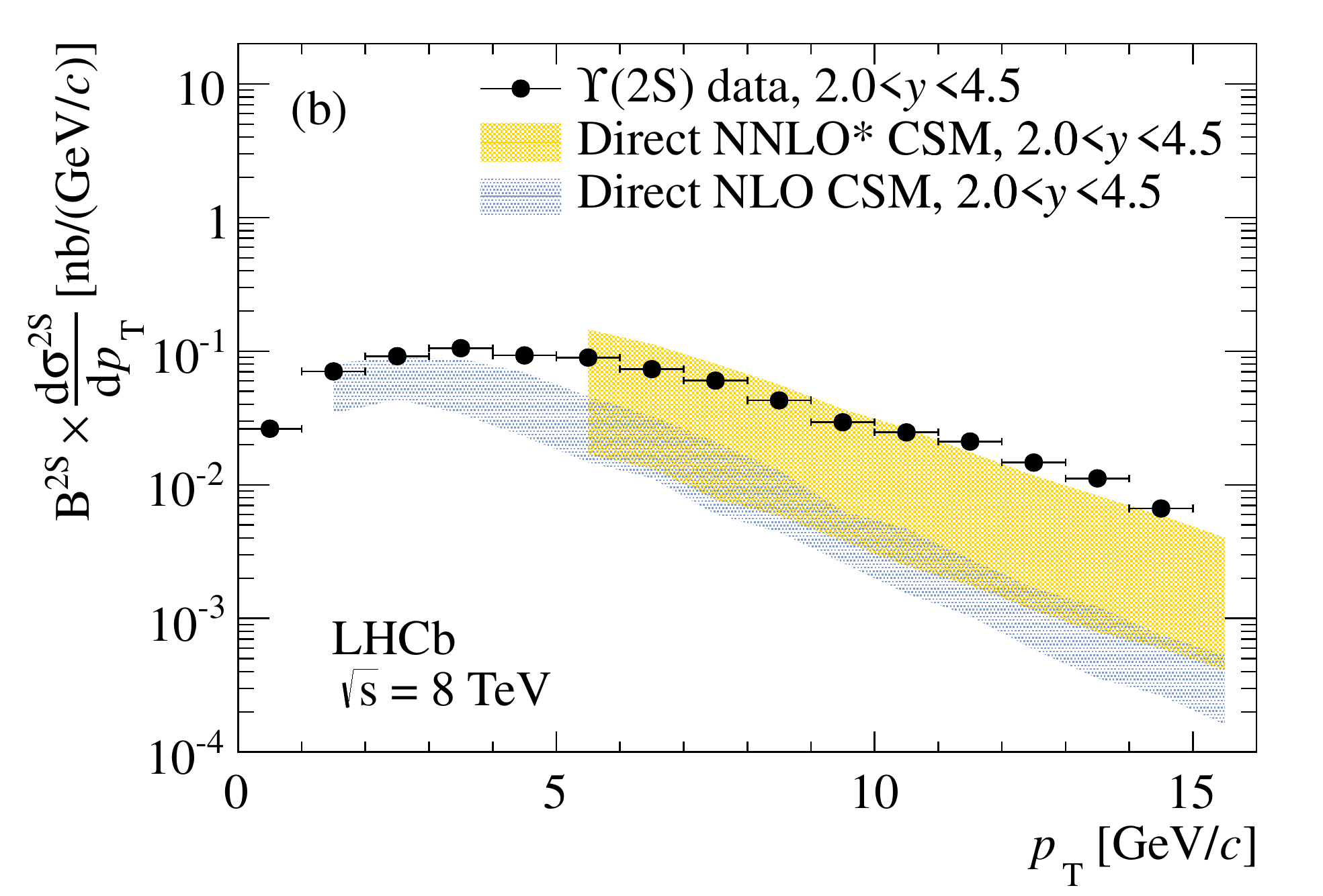}\\
  \includegraphics[width=7cm,clip]{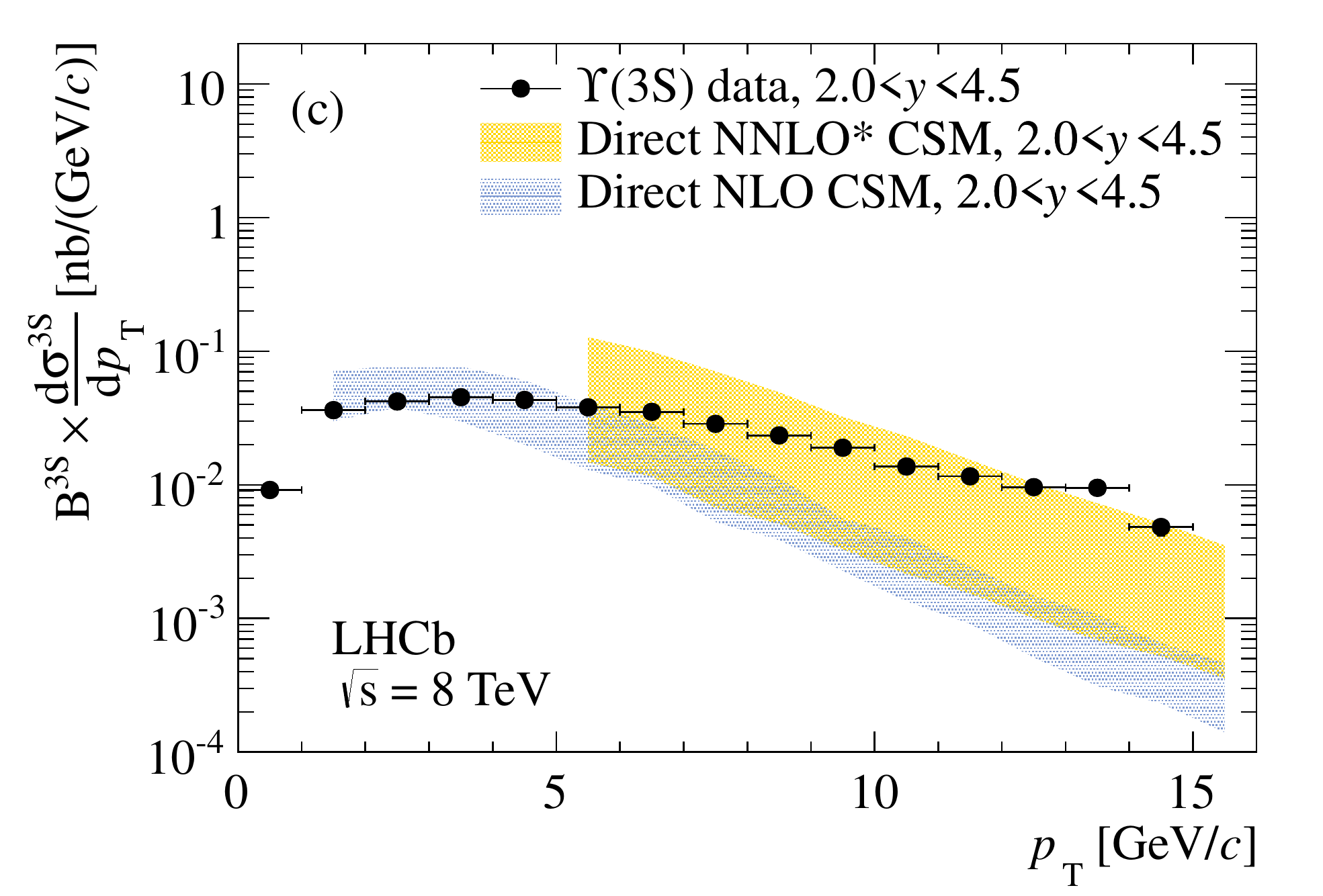}
 \caption{\small Comparison of the differential production cross-sections times dimuon branching fractions for (a) $\ones$, (b) $\twos$ 
  and (c) $\threes$ mesons  as a function of
   $\pt$   with direct production in an NNLO$^*$ CSM~\cite{Artoisenet:2008} (solid yellow) and an
   NLO CSM~\cite{Campbell:2007ws} (blue vertical shading) model. }\label{fig-7}
\end{figure}

In Fig.\ref{fig-7} the cross-sections times dimuon branching fractions for the 
three $\ups$ meson states are compared to the NNLO$^*$ CSM~\cite{Artoisenet:2008}  and an NLO CSM~\cite{Campbell:2007ws} predictions as a function of  $\pt$. The NNLO* CSM  provides a
reasonable description of the experimental data, particularly for the $\threes$ meson, which is expected to be less affected by feed-down. 
As for the prompt \jpsi\ meson production, the CSM at NLO underestimates the cross-section by an order of magnitude.
\section{Production of \psitwos\ mesons}
\label{sec-2}
\begin{figure}[h!]
\centering
\includegraphics[width=7cm,clip]{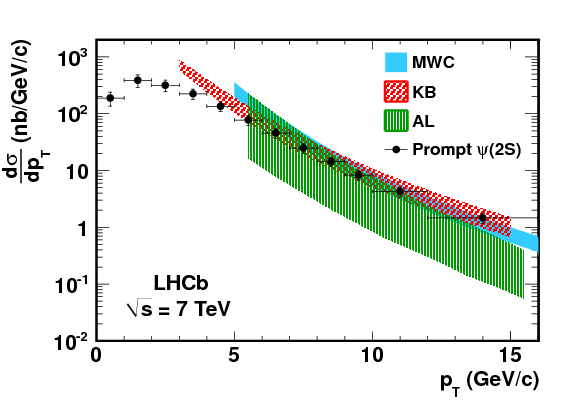}
\caption{\small  Differential production cross-section as a function of \pt\ for prompt \psitwos. The data are compared  with various theoretical models: MWC~\cite{Ma:2010jj} and KB~\cite{Butenschoen:2010rq} are NLO NRQCD calculations, while AL~\cite{Artoisenet:2008, lansberg:2009} is a CSM including the dominant NNLO terms }
\label{fig-8}       
\end{figure}

\begin{figure}[h!]
\centering
\includegraphics[width=7cm,clip]{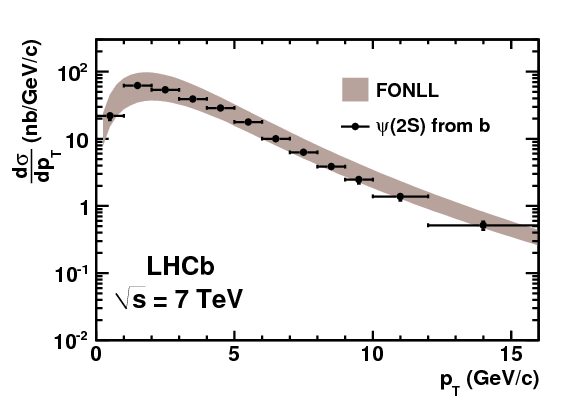}
\caption{\small Differential production cross-section as a function of \pt\ for \psitwos\ from $b$-hadron decays. The data are compared with FONLL calculations~\cite{FONLL,Cacciari}.  }
\label{fig-9}       
\end{figure}

The production of \psitwos\ mesons was studied by LHCb in $pp$ collisions at 7\tev~\cite{LHCb-PAPER-2011-045}. The \psitwos\ mesons were reconstructed in the decay channel $\psitwos\rightarrow\mu^{+}\mu^{-}$ and $\psitwos\rightarrow\jpsi\pi^{+}\pi^{-}$ and the results from the two channels were combined.  The \psitwos\ production from $b$-hadron decays is distinguished from promptly produced charmonium by exploiting the $b$-hadron finite decay time. The interest of measuring \psitwos\ production arises from the fact that no significant feed-down from excited charmonium states is expected, which facilitates the comparison with theoretical model. 
Figures~\ref{fig-8} and~\ref{fig-9} show the differential production cross-section as a function of \pt\ for prompt \psitwos\ and \psitwos\ from $b$-hadron-decays. 
The data are compared with various theoretical models: MWC~\cite{Ma:2010jj} and KB~\cite{Butenschoen:2010rq} are NLO NRQCD calculations, while AL~\cite{Artoisenet:2008, lansberg:2009} is a CSM including the dominant NNLO terms. The NLO NRQCD and FONLL~\cite{FONLL,Cacciari} calculations provide a good description of the  \psitwos\ prompt cross-section and of the cross-section of the  \psitwos\ from $b$-hadron decays.

\section{Double \jpsi\ production}
\label{sec-3}
\begin{figure}
\centering
\includegraphics[width=7cm,clip]{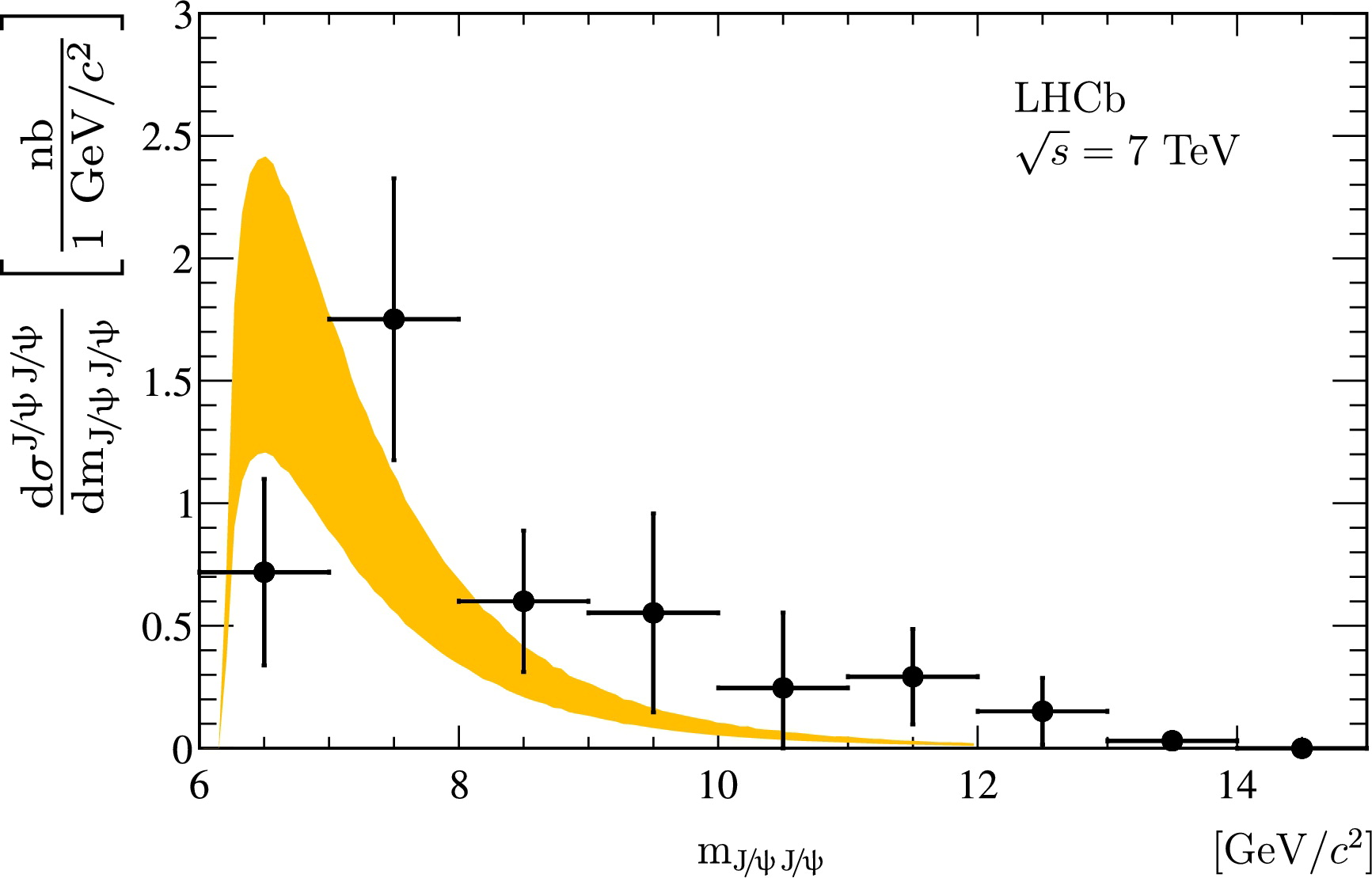}[htb]
\caption{\small Differential production cross-section for \jpsi\ pairs as a function of the invariant mass of the \jpsi\ pair system. The shaded area corresponds to the prediction by the model described in Ref.~\cite{Berezhnoy:2011xy}.  }
\label{fig-10}       
\end{figure}

Given the large charmonium production cross-section at the LHC, the question of multiple production naturally arises. This was studied by LHCb using an integrated luminosity of  about $37~\pbinv$ of $pp$ interactions at 7\tev~\cite{Aaij:2011yc}. Simultaneous production of two \jpsi\ mesons can be expected in single parton collisions via higher order $O(\alpha_s^4)$  gluon-gluon diagrams. This production could also be enhanced by double parton scattering (DPS) in which two gluons are extracted from a proton and the two \jpsi\ mesons are produced in two independent sub-processes. The cross-section for double \jpsi\ production in the fiducial range $2<y^{\jpsi}<4.5$ and $\pt^{\jpsi}<10\gevc$ is measured to be $\sigma^{\jpsi\jpsi}=5.1\pm1.0\pm1.1$nb~\cite{Aaij:2011yc}, where the first uncertainty is statistical and the second is systematic, in good agreement with  predictions from leading-order  QCD calculations~\cite{Berezhnoy:2011xy}. The suppression factor with respect to  single prompt \jpsi\ production  is $\sigma^{\jpsi\jpsi}/\sigma^{\jpsi}=(5.1\pm1.0\pm0.6^{+1.2}_{-1.0})\times 10^{-4}$, where the first uncertainty is statistical, the second systematic and the third accounts for the unknown \jpsi\ polarisation. Figure~\ref{fig-10} shows the differential production cross-section for \jpsi\ pairs as a function of the invariant mass of the \jpsi\ pair system. The data are compared with a theoretical prediction~\cite{Berezhnoy:2011xy}, which includes both direct production and feed-down from \psitwos\ decays and no contribution from DPS. Within the available statistics good agreement is observed. For double \jpsi\ production, the predictions from single and double parton scattering  are expected to be fairly close in magnitude~\cite{PhysRevD.86.034017}, but with somewhat different kinematic distributions for the \jpsi\ pairs. The analysis of the higher statistics already collected by LHCb will allow an in-depth study of the kinematic properties of double \jpsi\ production and of the different production models.

\section{Exclusive production of \jpsi\ and \psitwos\ mesons}
\label{sec-4}
\begin{figure}
\centering
\includegraphics[width=3cm,clip]{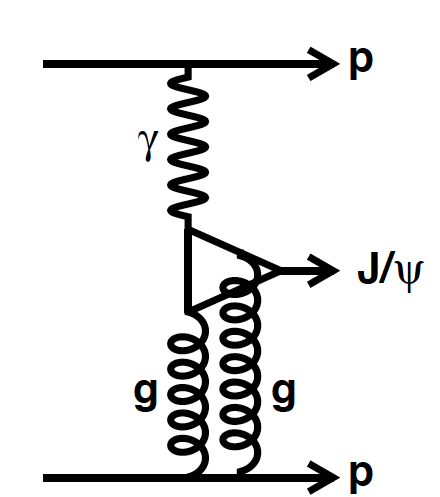}
\caption{\small  Feynman diagram contributing to exclusive \jpsi\ photoproduction.}
\label{fig-11}       
\end{figure}

Exclusive production of \jpsi\ and \psitwos\ mesons through processes such as the one displayed in Fig.\ref{fig-11} were observed by LHCb~\cite{Aaij:2013jxj} in the dimuon channel. LHCb is very well suited for this type of studies as it has access to high rapidities, with some sensitivity to backward tracks using the Vertex Locator (VELO), it operates at relatively low pileup and has sensitivity to low momentum and transverse momentum particles. Exclusively produced  \jpsi\ and \psitwos\ mesons into dimuons are characterised by a very distinct topology: their selection requires the presence of two reconstructed muons in the forward region and no other tracks and photons in the detector. An additional rapidity gap is obtained by also excluding tracks in the backward region using the VELO detector. Based on a data sample, corresponding to an integrated luminosity of about $36~\pbinv$ of $pp$ interactions, the cross-sections times branching fractions to two muons  for exclusive \jpsi\ and \psitwos\ with pseudorapidities between 2.0 and 4.5 is measured to be
$\sigma_{pp\to\jpsi(\to\mu^+\mu^-)} =307\pm 21\pm 36\,{\rm pb}$ and $\sigma_{pp\to\psitwos(\to\mu^+\mu^-)} =7.8\pm 1.3\pm 1.0\,{\rm pb}$, where the first  uncertainty is statistical and the second is systematic, in good agreement with theoretical predictions~\cite{HarlandLang:2009qe,PhysRevLett.92.142003,Motyka:2008ac,Goncalves:2011vf}. 
During the 2012 run, a new trigger was implemented to increase the rate of exclusive events, which makes use of ``upstream'' silicon sensors to veto any backward activity and of soft \pt\ cuts to gain sensitivity for hadronic final states. Data corresponding to an integrated luminosity of about $1.4~\mathrm{fb^{-1}}$ were already collected with such a trigger, which will allow LHCb to study the production of charmonium decaying into hadronic final states, the production of open charm and to perform a search for higher mass charmonium states.
\section{Production of \jpsi\ mesons in proton-lead collisions at $\sqrt{s_{\rm NN}}=5\tev$}
\label{sec-5}
\begin{figure}
\centering
\includegraphics[width=7cm,clip]{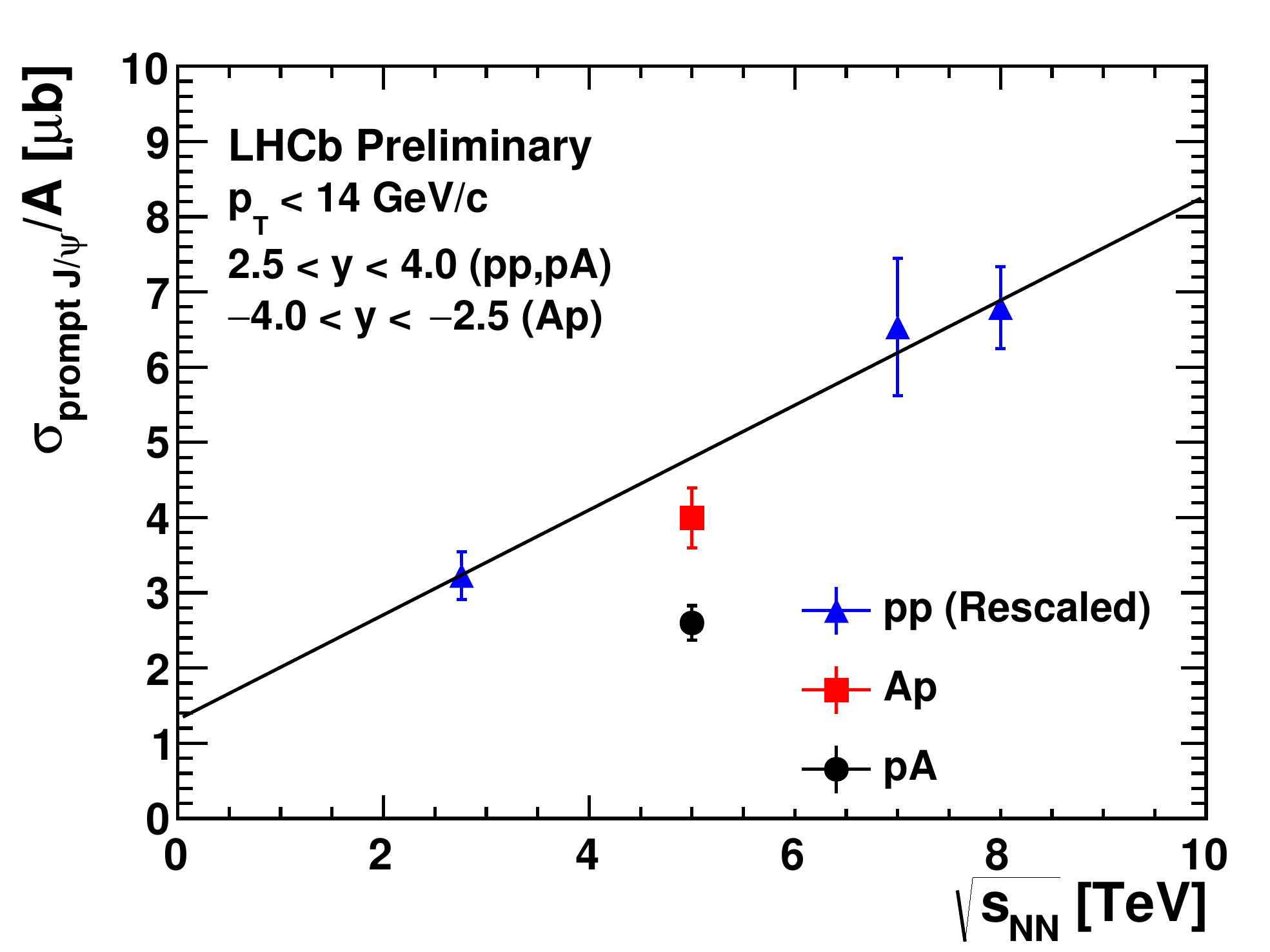}
\caption{\small Integrated cross-sections for prompt \jpsi\ production scaled by a factor of $1/A$ and rescaled to
the common rapidity range of the proton-nucleon system $2.5 < y< 4.0 $ for $pp$ and $pA$ collisions and $-4.0 < y< -2.5 $ for $Ap$ collisions as a function of $\sqrt{s_{\rm NN}}$}.
\label{fig-12}       
\end{figure}

\begin{figure}
\centering
\includegraphics[width=7cm,clip]{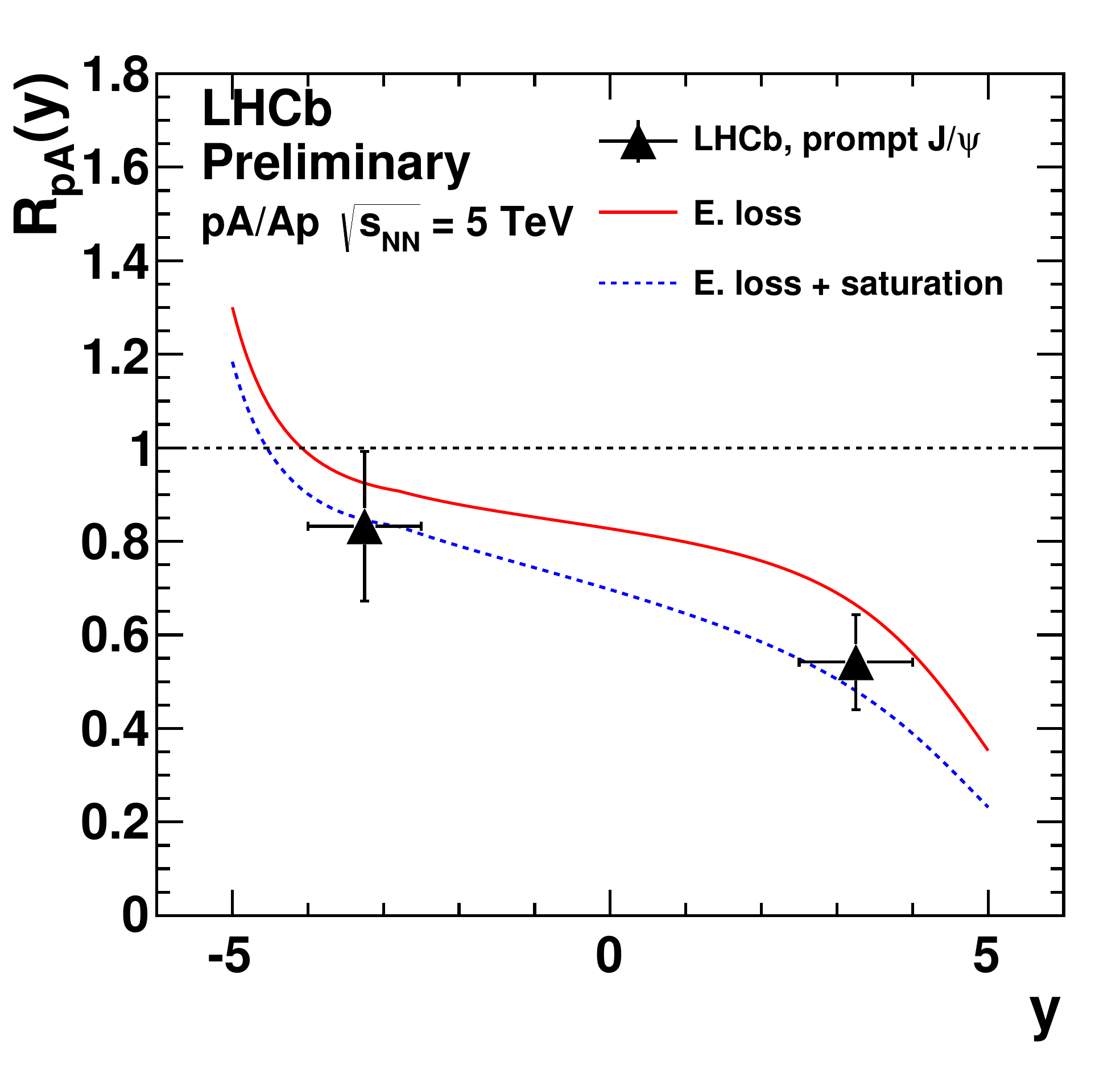}
\caption{\small Attenuation factor $R_{pA}$  compared to theoretical predictions~\cite{Arleo:2012rs}. The black triangles are LHCb measurements, the red solid line is the theoretical prediction based on parton energy loss effects, the blue dashed line takes additional saturation effects into account.}
\label{fig-13}       
\end{figure}

LHCb has recently presented preliminary results on production of \jpsi\ mesons in proton-lead collisions at $\sqrt{s_{\rm NN}}=5\tev$~\cite{LHCb-CONF-2013-008,Burkhard}. Data were collected with a proton beam energy of 4\tev\ and a lead beam energy of 1.58~\tev\ per nucleon, resulting in a centre-of-mass energy of the proton-nucleon system of 5~\tev. The directions of the proton and lead beams were swapped during data taking to produce both $pA$ and $Ap$ collisions. Data collected with the inverted beam directions ({\it i.e.}, in $Ap$ collision) allow LHCb to measure backward production. The proton-nucleon centre-of-mass system has a rapidity in the laboratory frame of +0.47 for $pA$ and -0.47 for $Ap$ collisions, resulting in a rapidity coverage in the laboratory frame ranging from about 1.5 to 4.0 for $pA$ collisions and from -5.0 to -2.5 for $Ap$ collisions. The analysis is based on data samples corresponding to integrated luminosities of 0.75~$\mathrm{nb^{-1}}$ of $pA$ and 0.30~$\mathrm{nb^{-1}}$ of $Ap$ collisions. The double-differential production is measured for prompt \jpsi\ and \fromb\ in bins of the kinematic variables $y$ and \pt. 
The measured cross-sections for prompt \jpsi\ production integrated over $y$ and \pt,  scaled by a factor of $1/A$ and rescaled to the common rapidity range of the proton-nucleon system $2.5 < y< 4.0 $ for $pp$ and $pA$ collisions and $-4.0 < y< -2.5 $ for $Ap$ collisions are shown is Fig.\ref{fig-12} as a function of $\sqrt{s_{\rm NN}}$. The results indicate that the \jpsi\ production cross-section is suppressed in proton-lead collisions, as expected from previous experiments; however such suppression is less enhanced  in the backward region. One can also derive the attenuation factor  $R_{pA}(y, \sqrt{s_{\rm NN}})=\frac{1}{A}\frac{\frac{{\rm d}\sigma_{pA}}{{\rm d}y}(y,\sqrt{s_{\rm NN})}}{\frac{{\rm d}\sigma_{pp}}{{\rm d}y}(y,\sqrt{s_{\rm NN})}}$ by using for the \jpsi\ production cross-section in $pp$ collisions at 5~\tev\ a linear interpolation of the published measurements at $\sqrt{s}=2.76$~\cite{LHCb-PAPER-2012-039}, 7~\cite{LHCb-PAPER-2011-003}, and $8\,\tev$~\cite{Aaij:2013yaa}. The result is shown in Fig.\ref{fig-13} and compared to theoretical predictions~\cite{Arleo:2012rs}. Within the available statistics good agreement is observed.


\section{Conclusion}
\label{sec-6}
LHCb has performed a wealth of measurements on quarkonium production, and many more are coming! A few selected ones were briefly illustrated in this report. 

Measurements of charmonium~\cite{LHCb-PAPER-2011-003,LHCb-PAPER-2012-039,LHCb-PAPER-2011-045} and bottomonium~\cite{LHCB-PAPER-2011-036,Aaij:2013yaa}  production were performed at various centre-of-mass energies. They allow an in depth comparison with theoretical models. A simple CSM is disfavoured by the data, while a combination of CS and CO, as implemented in the NRQCD formalism, or CS improved by QCD corrections, provide a good description of prompt quarkonium production.  Charmonium production from $b$-hadron decays is very well reproduced by FONLL calculations.
Some other studies were illustrated in this report, such as charmonium exclusive production~\cite{Aaij:2013jxj}, double \jpsi\ production~\cite{Aaij:2011yc}, as well as first preliminary results on \jpsi\ production in proton-lead collisions at $\sqrt{s_{\rm NN}}=5\tev$~\cite{LHCb-CONF-2013-008}. Many of these results are based on the analysis of a partial dataset and will gain in precision with the analysis of the full data sample available to LHCb.
\section{Acknowledgments}
I would like to thank the Organizers of LHCP 2013 for their kind invitation, my
LHCb colleagues for providing the material discussed here and, in particular, Giulia Manca, Vanya Belyaev and Burkhard Schmidt for
their careful reading of this article.


%
\bibliography{Mybib}
%
%
%

\end{document}